\documentclass{aa} 

\usepackage{amsmath}
\usepackage{tablefootnote}
\usepackage{bm}
\usepackage{xcolor}
\usepackage[normalem]{ulem}

\usepackage{graphicx}
\usepackage{txfonts}
\usepackage[colorlinks,linkcolor=blue,citecolor=blue,linktocpage=true,breaklinks, 
plainpages=false,urlcolor=blue]{hyperref}

\begin{document}

   \title{Spatial resolution effects on the solar open flux estimates}

   \titlerunning{Spatial resolution effects on the Solar open flux estimates}
   \authorrunning{}

   \author{I. Mili\'{c}
          \inst{1,2}
          \and
          R. Centeno
          \inst{3}
          \and 
          X. Sun
          \inst{4}
          \and 
          M. Rempel
          \inst{3}
          J. de la Cruz Rodr\'iguez
          \inst{5}
          }

   \institute{Institute for Solar Physics (KIS), Sch\"{o}nek Str. 6, 79111 Freiburg, Germany\\
   \email{milic@leibniz-kis.de}
   \and
   Faculty of Mathematics, University of Belgrade, Studentski Trg 16, 11000 Belgrade, Serbia
   \and
   High Altitude Observatory (NCAR), 3080 Center Green Dr., Boulder, CO, 80301, USA
   \and
   Institute for Astronomy, University of Hawai`i at M\={a}noa, 34 Ohia Ku Street, Pukalani, HI 96768, USA
   \and
   Institute for Solar Physics, Dept. of Astronomy, Stockholm University, AlbaNova University Centre, SE-106 91 Stockholm, Sweden
   }

   \date{Received ; accepted }

 
  \abstract
   {Spectropolarimetric observations used to infer the solar magnetic fields are obtained with a limited spatial resolution. The effects of this limited resolution on the inference of the open flux over the observed region have not been extensively studied. } 
   {We aim to characterize the biases that arise in the inference of the mean flux density by performing an end-to-end study that involves the generation of synthetic data, its interpretation (inversion), and a comparison of the results with the original model.}
   {We synthesized polarized spectra of the two magnetically sensitive lines of neutral iron around 630\,nm from a state-of-the-art numerical simulation of the solar photosphere. We then performed data degradation to simulate the effect of the telescope with a limited angular resolution and interpreted (inverted) the data using a Milne-Eddington spectropolarimetric inversion code. We then studied the dependence of the inferred parameters on the telescope resolution.}
   {The results show a significant decrease in the mean magnetic flux density ---related to the open flux observed at the disk center--- with decreasing telescope resolution. The original net magnetic field flux is fully resolved by a 1m telescope, but a 20\,cm aperture telescope yields a 30\% smaller value. Even in the fully resolved case, the result is still biased due to the corrugation of the photospheric surface.}
   {Even the spatially averaged quantities, such as the open magnetic flux in the observed region, are underestimated when the magnetic structures are unresolved. The reason for this is the presence of nonlinearities in the magnetic field inference process. This effect might have implications for the modeling of large-scale solar magnetic fields; for example, those corresponding to the coronal holes, or the polar magnetic fields, which are relevant to our understanding of the solar cycle.}


   \maketitle
   
%

\section{Introduction}

The physical parameters in the solar photosphere are generally inferred from the intensity and polarization of the observed light (i.e., the polarized spectrum) at the wavelengths corresponding to spectral line transitions. Arguably the most interesting parameter is the magnetic field vector, the measurements of which are essential for our understanding of the active regions, flux emergence, magnetic reconnection, and various other aspects of solar activity and dynamics. Notably, the photospheric magnetic field is used as a boundary condition for the so-called magnetic field extrapolation techniques \citep{Regnier2013}, which allow us to model the magnetic field of the outer solar atmosphere and the heliosphere. Specifically, the extrapolations of the magnetic fields from the regions that correspond to the coronal holes into the heliosphere yield a significant mismatch with the values obtained from in situ measurements \citep[the open flux problem; see][]{Linker2017}. A possible way to resolve this problem is related to the fact that our current photospheric diagnostics underestimate the open magnetic flux. 

Reasons for a mismatch between a ``true'' value, which is, strictly speaking, never known to us, and the inferred one can be numerous. 
First, simplified spectral line formation and polarization models used for magnetic field diagnostics introduce systematic errors that are non-trivial to find and eliminate. The limited spatial, spectral, and temporal resolution of our observations biases the results of our diagnostics. Non-negligible photon noise results in significantly different errors in the inference of line-of-sight and transversal components of the field. Finally, even a perfect inference yields two equally valid orientations of the magnetic field, with different physical implications. Away from the disk center, this results in different vectors in the local reference frame, different radial components of the magnetic field, and, therefore, different open fluxes.

A complete assessment of all of these factors is only possible if we simulate the whole measurement process and compare the output to the input values of the parameters used to calculate the synthetic observations. \citet{LekaBarnes2012} performed such a study using examples of synthetic and observed data and found differences between the original and inferred mean unsigned flux density and the field inclination. Furthermore, \citet{2016A&A...594A.103D} and \citet{2016A&A...593A..93D} studied the effect of telescope PSF straylight in the retrieved inter-network magnetic field vector by means of 2D spatially coupled inversions of Hinode data. These latter authors showed that the mean magnetic field at an optical depth of unity is approximately 130~G. At higher layers, the field strength is lower and the field is more horizontal. Recently, an end-to-end study was performed for the GONG instrument \citep{GONG} and is described in a series of papers by \citet{Plowman1, Plowman2, Plowman3}, who pointed out numerous biases in the measurement process. Studies like these are important because they allow us to modify the input models and to therefore better understand the inference biases. 

In the following analysis, we focus on the effects of the limited spatial (i.e., angular) resolution of the telescope on the magnetic field inference at the center of the solar disk. In this physical scenario, the line-of-sight component of the field coincides with the radial direction. That is, our inference (inversion) directly yields the radial component of the field. While a poorer spatial resolution clearly results in a loss of small-scale details, one would not expect a change in the net inferred radial magnetic field, that is, averaged over the whole field of view (FoV). However, because the magnetic field inference is a nonlinear process, we find the opposite outcome. In the following section, we show how we synthesize  the observational data and the magnetic field inference. We follow up with section 3, where we describe our magnetic field inversion and a comparison between the original and inferred magnetic fields. Section 4 presents an illustration of this effect on the real-life observed data, where SST/CRISP \citep{SST, CRISP} observations are used as the ground truth. In section 5, we present our conclusions and plans for future work.

\section{Data preparation and inversion}

\subsection{Synthetic spectra calculation}

\begin{figure*}
\includegraphics[width=0.99\textwidth]{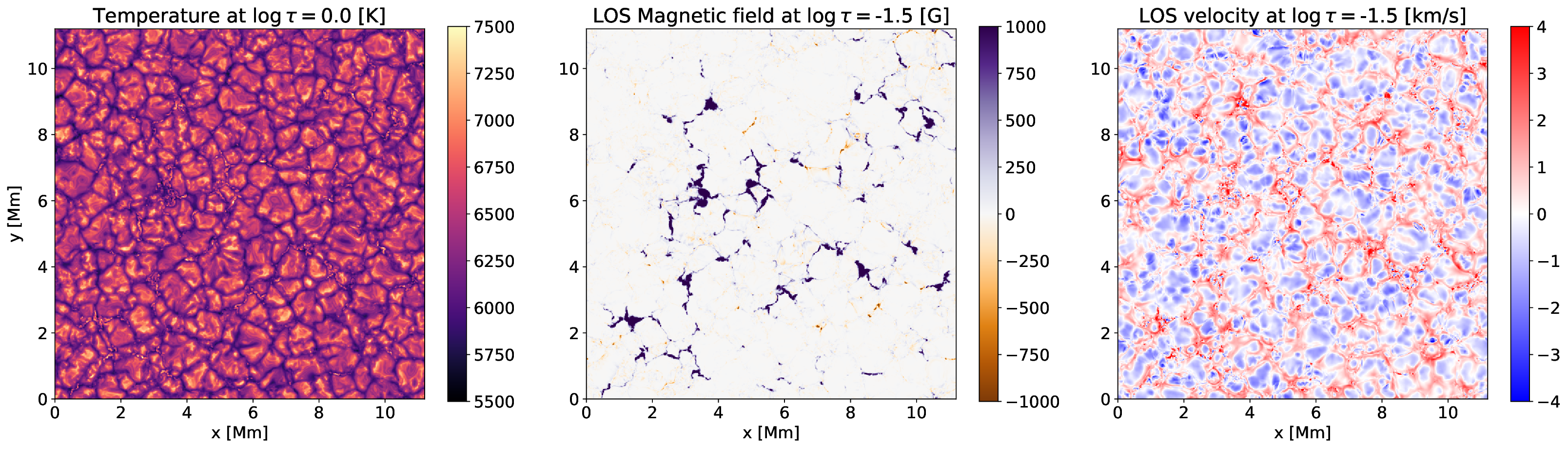}
\caption{Temperature at the photosphere ($\log\tau=0$), and the line-of-sight magnetic field and velocity at ($\log\tau=-1$) in the simulated solar photosphere used in this study.}
\label{simulation}
\end{figure*}

In this study, we use a state-of-the-art radiative-magnetohydrodynamic (RMHD) simulation of the solar atmosphere performed with the MURaM code \citep{Vogler2005, Rempel2014}. Three-dimensional photospheric models obtained using MURaM have been shown to accurately reproduce the observed properties of the solar atmosphere \citep[e.g.,][]{SanjaHinode}. We use a single snapshot based on a small-scale dynamo simulation presented in \citet{Rempel2014}, which considers a computational box with horizontal extent of 24.576 $\times$ 24.576\,Mm (1536 $\times$ 1536 cell point with 16\,km spacing) and vertical extent of 8.192\, Mm (512 points with 16 km spacing). A uniform field of 30 G was added to the small-scale dynamo solution with zero net flux and the simulation was evolved for an additional 6 hours to allow for the formation of a magnetic network. The last 15 minutes were evolved with nongray radiative transfer. In the layers corresponding to the solar photosphere, the magnetic field forms a complicated, plage-like magnetic structure (see Fig.\,\ref{simulation}) with patches of kG fields. While the spatial ($x,y$) structure of the field is changing, the mean vertical magnetic field is constant with height, that is, the vertical flux is conserved as a consequence of periodic boundary conditions in the horizontal directions. If we consider this atmosphere as a ground truth, an ideal diagnostic method would retrieve a mean flux density equal to 30\,G.  This mean flux density corresponds to an open flux of $\approx 1.8\times10^{20}$\,Mx. In the context of this study, we refer to this flux as ``open'', even though the open flux is observationally typically defined over much larger areas. 

We calculated the spectrum of the two magnetically sensitive neutral iron lines around 630\,nm using the SIR code \citep{SIR}. These are the spectral lines observed by the HINODE SOT/SP \citep{Hinode} instrument and are a common choice for thermodynamic and magnetic diagnostics of the solar photosphere. As we are primarily concerned with the magnetic field, the synthesis is carried assuming local thermodynamic equilibrium (LTE). Strictly speaking, iron is overionized in the photosphere and precise modeling of the iron lines requires a non-LTE approach \citep[e.g.,][]{Thevenin99, 2001ApJ...550..970S,SmithaNLTE2}.

The calculation of the spectra yields a single polarized spectrum for each pixel, sampled at the range from 630.1 to 630.3\,nm with a 1\,pm step size. For a given model of the solar atmosphere, this is the spectrum we would measure provided an instrument with infinite spatial (and spectral) resolution. When observed with a telescope, spatial distribution of the incoming polarized intensity gets convolved with the point spread function (PSF) of the optical system, which is often approximated by the diffraction pattern of the entrance aperture. We assume an ideal, Airy disk PSF (diffraction pattern of a circular aperture) for four different diameters, namely 0.1, 0.2, 0.5, and 1.0\,m, and convolve synthetic monochromatic images of each Stokes parameter at each wavelength with these PSFs to get four datasets, corresponding to different spatial resolutions. For reference, a 1\,m circular aperture at this wavelength results in the resolution of 0.16 arcseconds, which corresponds to approximately 120\,km on the solar surface, or eight times larger than the resolution pixel. Therefore, we perform a 2$\times$2 binning of the convolved images to ease the manipulation and reduce inversion time. We note that the Airy disk PSF alone is not enough to fully reproduce the spatial distribution of the emitted radiation. \citet{SanjaHinode} demonstrated that additional instrumental effects, such as defocus, binning, and so on, have to be considered. Therefore, the effects described here would be slightly different for examples of specific telescopes and instruments. 

\subsection{Inversions}

Inversion is the process of fitting the observed Stokes spectrum with a physically motivated model. The model parameters are the physical quantities (in this specific case, magnetic field vector) that we want to infer from the observed Stokes vector. The most realistic model is a fully stratified atmosphere where temperature, pressure, magnetic field, and velocity are allowed to freely vary with height. However, fitting such a model to the data is a fairly difficult task \citep[see the review by][for an in-depth description of spectropolarimetric inversions]{LRSP}, and simplifications are often needed. A robust trade-off between realism and simplicity is the assumption of a Milne-Eddington atmosphere \citep[M-E, e.g.][]{dtibook}. In this case, spectral line formation is modeled by several parameters that are constant with depth (magnetic field vector, line-of-sight velocity, line strength, Doppler broadening, and damping) and a line source function that varies linearly with optical depth. The M-E atmosphere does not explicitly assume LTE, but it is most often used to model the LTE lines formed in the photosphere, as there the source function can be reasonably well approximated with a linear function. Codes based on the M-E model are widely used in the data-inversion pipelines, such as VFISV \citep{VFISV}, used for the inversion of SDO/HMI observations, MERLIN \citep{MERLIN}, used for Hinode SOT/SP data, and MILOS \citep{2007A&A...462.1137O}, used for Solar Orbiter data.

Here, we chose a recently developed M-E inversion code, PyMilne\footnote{\url{https://github.com/jaimedelacruz/pyMilne}} \citep{PyMilne}. The code is compiled in C++, usable from Python, and makes full use of OpenMP capabilities, thus enabling very fast inversion times even on personal computers (hundreds to thousands of spectra per second). The use of a M-E inversion scheme calls for a discussion on the vertical and horizontal dependence of the magnetic field, that is, a consideration of the line-formation height and the magnetic field filling factor. 

The magnetic field in the solar atmosphere undoubtedly varies with depth, even on scales comparable to spectral line formation regions. As the M-E model assumes a constant magnetic field, it is not straightforward to relate the inferred magnetic field vector to a specific atmospheric depth. A comprehensive study by \citet{Fistro2014} showed that the M-E inversions retrieve a weighted mean of the magnetic field over the line formation depths, where the weighting function is the so-called response function \citep{RealRF}. If we want to relate inferred magnetic field over an extended FoV to a specific height, or even optical depth, we are introducing some bias as we probe the field at different depths in different pixels. We show that this effect leads to a disagreement in the inference of the net flux even in the absence of the telescope PSF (section \ref{depthtau}). 

Another aspect is related to the limited angular resolution. In principle, the magnetic field in the solar photosphere might be structured on very small scales, that is, certainly smaller than the pixel size. The spectrum of the observed pixel can then be modeled as a weighted combination of a magnetic and nonmagnetic atmosphere, where the weighting is typically described by the so-called filling factor. The nonmagnetic component is often referred to as stray light even though it does not necessarily physically describe the stray light in the telescope. Recently, we assessed the role of the filling factor in the polar Hinode-like observations \citep{paperI} and found that M-E inversions that use the filling factor still yield very large errors when inferring intrinsic magnetic field properties. Additionally, the filling factor can result in a severe overfitting of the observed spectra, and potentially dubious interpretation of the results (e.g., pixels with very small polarization signals can still yield strong magnetic fields). Therefore, in the following analysis, we assume that the observed pixel is completely permeated by the magnetic field, that is, that the filling factor is equal to unity. We note that PyMilne allows the use of spatially coupled inversion \citep[discussed in][]{PyMilne}, which, in principle, can also include the known telescope PSF \citep[similar to][]{vanNoort2012}, which is a feature we will explore in a follow-up paper to this one.

\section{Inversion of synthetic observations}\label{sec:inversions}

We inverted the original synthetic dataset (1536 $\times$ 1536 polarized spectra), as well as the four degraded datasets (768 $\times$ 768 spectra each) that correspond to telescope apertures of 1\,m, 0.5\,m, 0.2\,m, and 0.1\,m (assuming ideal, clear-aperture telescopes). Each of the inversions yields a single value for the magnetic field vector for each pixel. This allows us to analyze the inferred vertical component of the magnetic field and therefore the mean magnetic flux density. The latter multiplied by the surface area of the FoV yields the open flux. We note that we use ``inferred'' here to emphasize that these values come from the inversion of the polarized spectra, given the assumed model, and are different from the true values. 

\begin{figure*}
\includegraphics[width=0.99\textwidth]{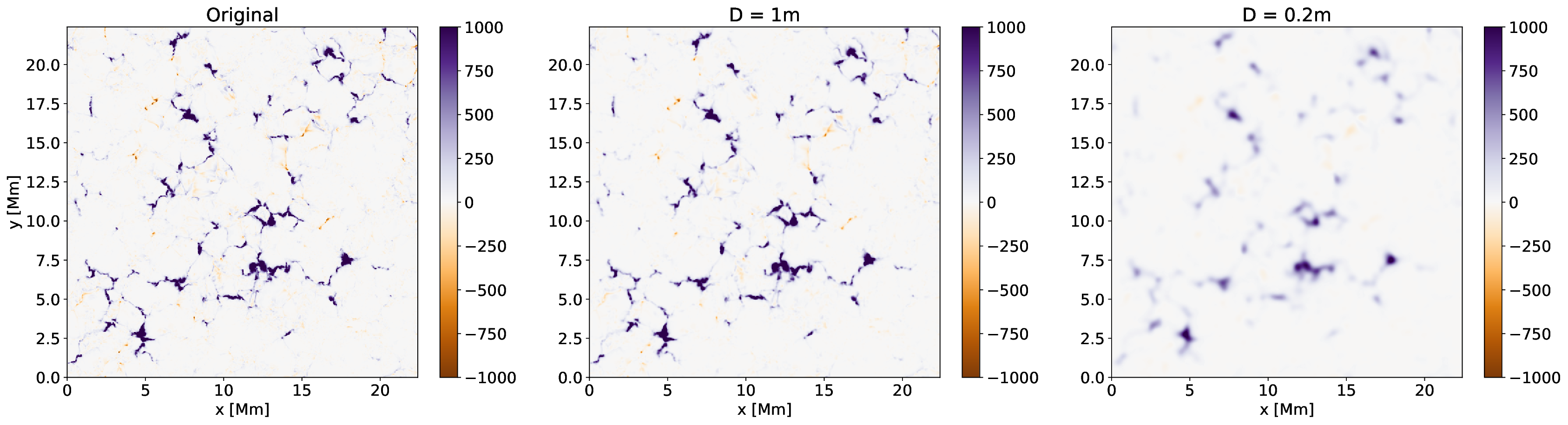}
\caption{Line-of-sight magnetic field as inferred from the synthetic observations using Py Milne. Left: Original dataset. Middle: Dataset degraded according to the PSF of a 1m telescope. Right: Same as the middle panel but for a 0.2\,m telescope. All three plots are in units of Gauss. The mean flux density for these three cases is $32.8$\,G, $32.6\,$G, and $22.9\,$G,  respectively.}
\label{bmaps}
\end{figure*}

Figure \ref{bmaps} shows the spatial distribution of the inferred line-of-sight magnetic field for original data, and the spatial distributions corresponding to 1\,m aperture and 0.2\,m aperture size. Visually, there is no significant difference between the first two, while the third map shows significantly weaker and more poorly resolved magnetic fields. On
the other hand, the inferred mean flux density for the three cases is equal to 32.8\,G, 32.6\,G, and 22.9\,G, respectively. Another quantity that is often analyzed is the mean unsigned flux density, which for these three
cases is equal to 48.9\,G, 41.9\,G, and 25.9\,G, respectively. 

It is clear that, due to the apparent cancellations, the decrease in the telescope resolution causes us to see less magnetic field overall. This is reflected in the decrease in the mean unsigned flux density with decreasing spatial resolution, as discussed  by, for example, \citet{AM2011}. In their well-known Figure 3, these latter authors compare a large number of observational studies of the quiet Sun from multiple authors and show that the mean unsigned flux density decreases monotonically with decreasing spatial resolution. Here, the mean unsigned flux density is the average value of $|B_z|$ over the FoV, where $B_z$ is the vertical component of the magnetic field. The mean magnetic flux density, on the other hand, is equal to the average of $B_z$, and, when multiplied by the area of the FoV, gives us the total open flux. The open flux is proportional to the number of magnetic field lines that extend vertically outside of the observed region. We denote mean magnetic flux density with $\langle B_z \rangle$ and mean unsigned flux density with $\langle |B_z| \rangle$.

\begin{figure}
\includegraphics[width=0.49\textwidth]{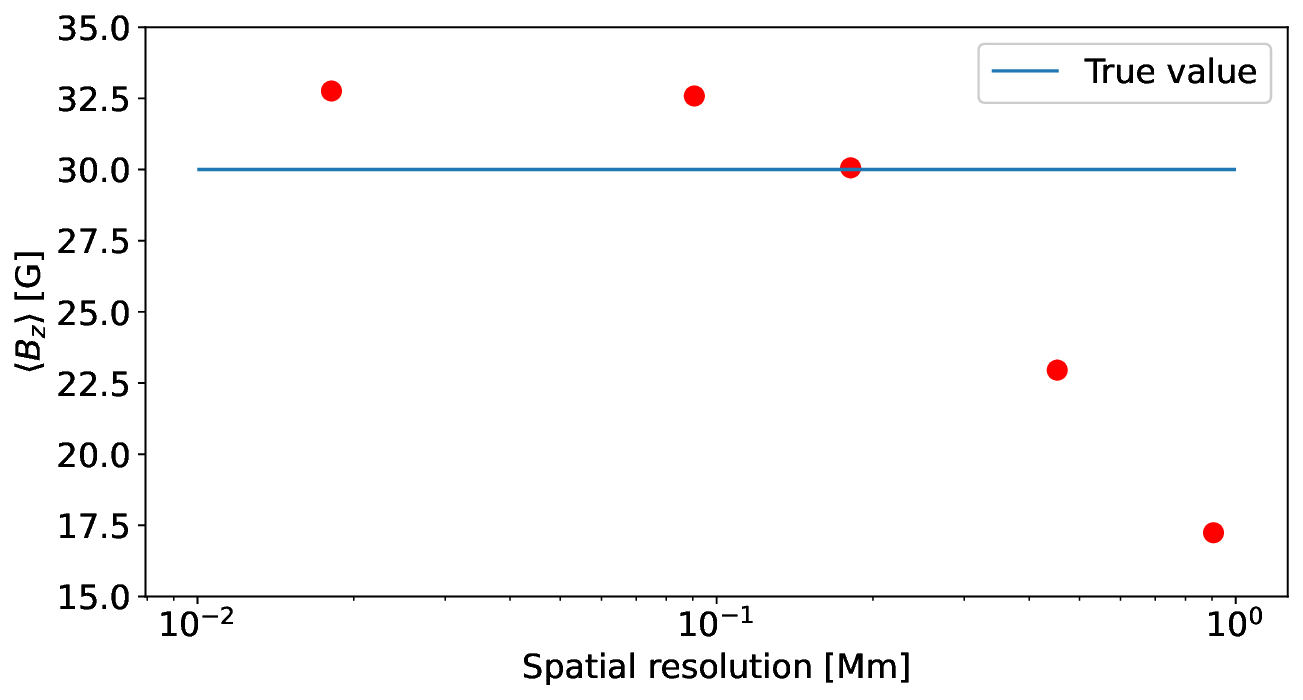}
\includegraphics[width=0.49\textwidth]{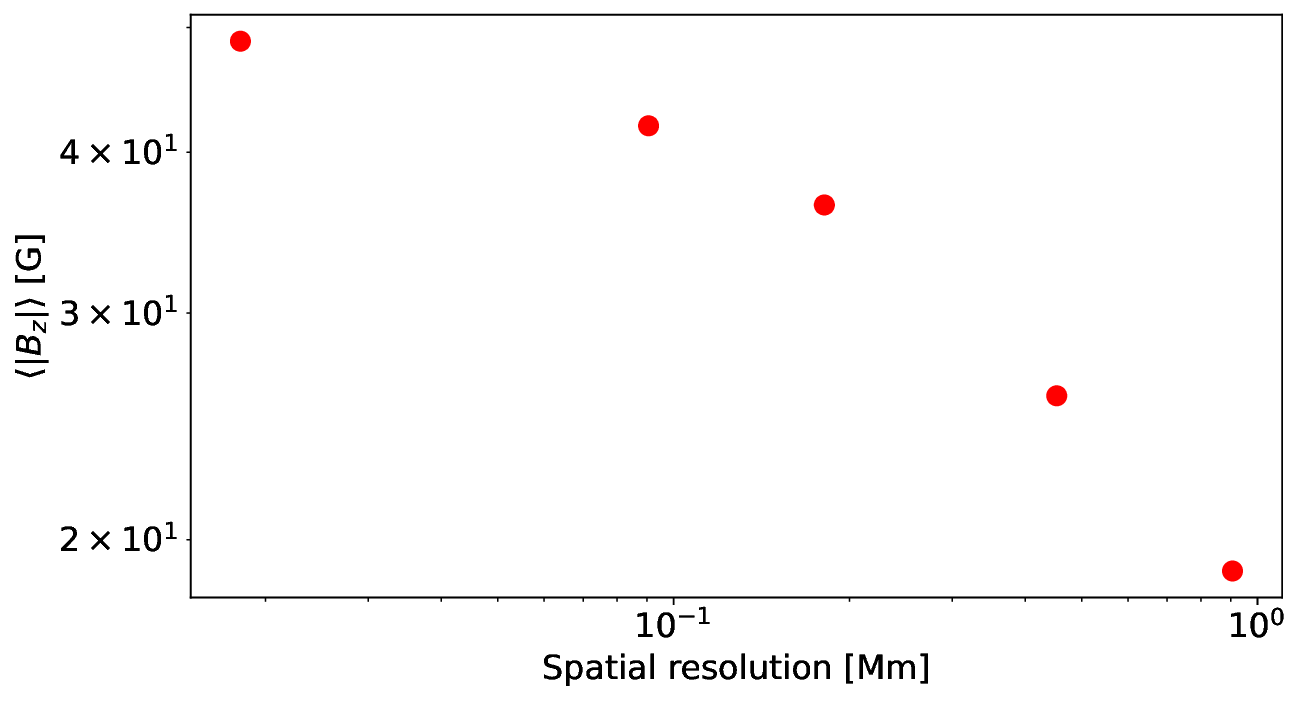}
\caption{Effects of the telescope resolution on the average of the inferred magnetic quantities. Top: The inferred mean flux density. The true value from the model atmosphere is 30\,Gauss. Bottom: Same, but for the mean unsigned flux density.}
\label{fluxes}
\end{figure}

\begin{figure}
\includegraphics[width=0.49\textwidth]{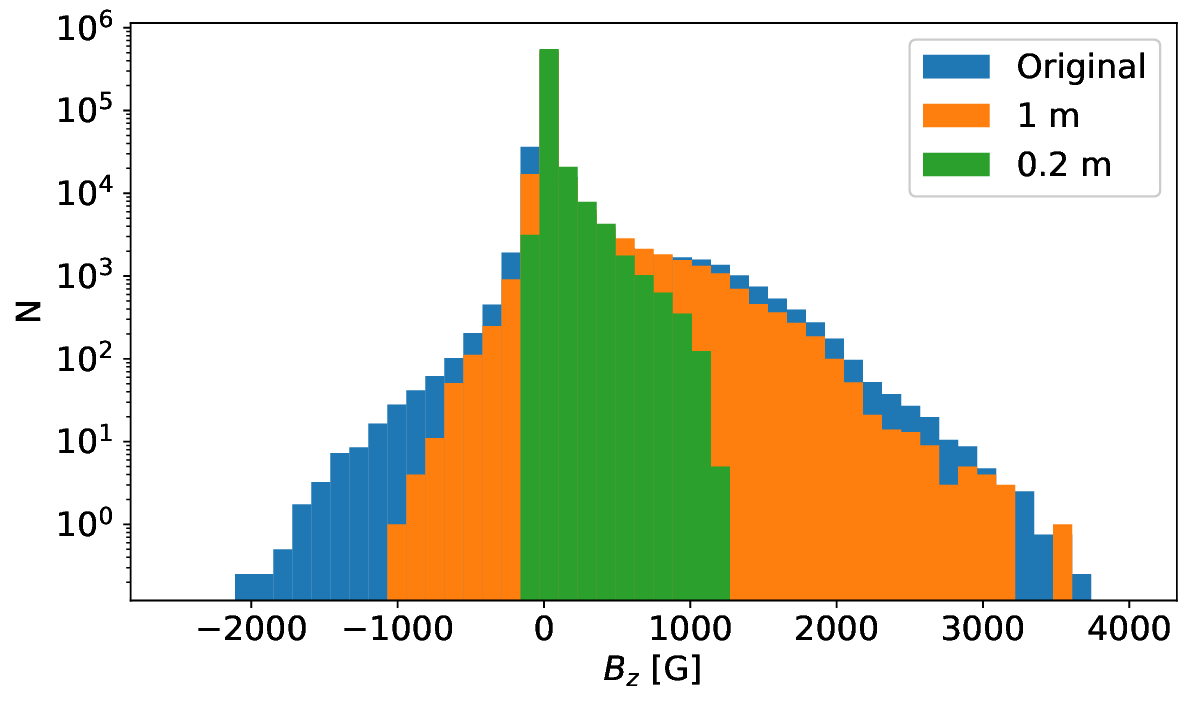}
\caption{Distribution of the inferred magnetic flux density for original synthetic observations and the observations degraded according to the 1m and the 0.2m telescope PSF.}
\label{Bzdist}
\end{figure}

While it is clear that the mean unsigned flux density must decrease with telescope resolution because of the cancellation of the small-scale magnetic features, intuitively we expect the mean flux density to be constant. The reason for this is that, while poor telescope resolution might result in the cancellation of polarization signals, we expect to see approximately the same number of open field lines, even after the spatial averaging. Our results (Fig.\,\ref{fluxes}, top) reveal that this is not the case. $\langle B_z \rangle$ in the original atmospheric cube is 30\,G. The value inferred from the undegraded synthetic spectra is slightly higher: 32.8\,G. This value does not change if we convolve our data with a PSF that corresponds to the 1m telescope (32.6\,G). However, convolution with the PSF of a 0.2m telescope yields a $30\%$ decrease (22.9\,G). The upper panel of figure \ref{fluxes} shows that, below the spatial resolution of $\approx 100$\,km (1m telescope), the estimated $\langle B_z \rangle$ quickly decreases down to almost half of its value, which is reached at the 10\,cm aperture. For completeness, we also show (Fig.\,\ref{fluxes}, bottom) the relationship between the $\langle |B_z| \rangle$ and the telescope resolution \citep[corresponding to Fig.\,3 of ][]{AM2011}, in the lower panel figure\,\ref{fluxes}. We note that the latter is the log-log scale in order to make it comparable with the results of \citet{AM2011}. In addition to the mean of the vertical magnetic flux density, the whole inferred distribution significantly changes with the telescope resolution. Figure\,\ref{Bzdist} shows the changes in the inferred distribution of $B_z$ with the telescope size. While the distribution of data degraded with the 1m telescope PSF only departs from the original distribution in the most extreme values, the degradation with the 0.2m telescope PSF completely removes magnetic fields stronger than 1 kiloGauss. For reference, the standard deviations for the original and 1m cases are 179 and 159 Gauss, respectively, while the application of the 0.2m PSF decreases this all the way down to 79\,Gauss.

\subsection{Inversion nonlinearities}

The reason for the decrease in the inferred mean flux density lies in the nonlinearity of radiative transfer processes that relate the atmospheric parameters to the emergent Stokes vector and, conversely, in the inversion process. The Stokes vector $\hat{I}_\lambda(x,y)$ emerging from a 1D atmosphere at the location $(x,y)$ is a result of a nonlinear mapping that involves calculation of the emission and absorption terms, as well as the solution of the polarized radiative transfer equation \citep[see e.g.,][]{dtibook}. We denote this mapping by a functional $\mathcal{F}$:
\begin{equation}
    \hat{I}^0_{\lambda}(x,y) = \mathcal{F}[\vec{B}(x,y,z), T(x,y,z), \vec{v}(x,y,z)] = \mathcal{F}[\vec{M}^0(x,y)].
\end{equation}
Hereafter, we use $\vec{M}^0(x,y)$ to refer to all the physical parameters relevant to the polarized line formation, absorbing their depth-dependence in the vector $\vec{M}^0$. In the detector, we measure the Stokes vector convolved with the telescope PSF:
\begin{equation}
    \hat{I}_\lambda(x,y) = \int \int \hat{I}^{0}_\lambda(x',y') PSF(x'-x,y'-y) dx'dy',
\end{equation}
which translates to
\begin{equation}
    \hat{I}_\lambda(x,y) = \int \int \mathcal{F}[\vec{M}^0(x,y)] PSF(x'-x,y'-y) dx'dy'.
    \label{nonlinear}
\end{equation}
The estimated physical parameters are inferred (e.g., using an M-E inversion) by performing an inverse of the functional $\mathcal{F}$:
\begin{equation}
\vec{M}(x,y) = \mathcal{F}^{-1}[\hat{I}_\lambda(x,y)].
\end{equation}
Therefore, the relationship between the measured physical parameters and the original ones is 
\begin{equation}
\vec{M}(x,y) = \mathcal{F}^{-1}[PSF \circledast \mathcal{F}[M^0(x,y)]]
,\end{equation}
where $\circledast$ denotes convolution in the $x,y$ plane. While the convolution is a linear operator, the radiative transfer and the inversion process are not, and so the two will not commute. Therefore, we cannot say that the inferred physical parameters in the presence of the telescope PSF are simply the original parameters convolved by the telescope PSF. That is, a spatially averaged magnetic field obtained from inversion is not necessarily identical to the spatial average of the original magnetic field. This has already been pointed out, for example, by \citet{Han2011} for the case of spectral line intensity, and recently by \citet{Plowman2} and \citet{paperI} for the case of the photospheric magnetic field vector. Additional biases are caused by the imperfections in the inverse operator $\mathcal{F}^{-1}$ which cannot account for all of the complexities in the forward model ${\mathcal F}$, and is often a very rough simplification of reality.

In the case of simple photospheric magnetic field diagnostics, this bias can be illustrated by considering a weak field approximation \citep[WFA; e.g.,][]{dtibook}. Here, the circular polarization in the line is:
\begin{equation}
    V = k B_{\rm los} \frac{dI}{d\lambda}, 
\end{equation}
where the constant $k$ depends on the spectral line in question. For the disk-center case: $B_{\rm los} = B_z$. ${dI}/{d\lambda}$ depends on the depth gradient of the source function, which is generally smaller in intergranular lanes than in granules. On the other hand, intergranular lanes typically harbor stronger fields than granules. Considering four neighboring pixels, that is, one magnetic intergranular pixel, and three nonmagnetic granular pixels, and assuming that the intergranular pixel is permeated by $B_z = 1000$\,G field, but has a two-times smaller value of ${dI}/{d\lambda}$ than the granular one: if we can resolve pixels individually and infer their magnetic fields, we will estimate the mean flux density to be $\langle B_z \rangle = {1000 {\rm G}}/{4} = 250$\,G. Now, if we cannot resolve individual pixels, $V$ and $dI/d\lambda$ will follow from:
\begin{align}
    \left (\frac{dI}{d\lambda} \right )_g &= 2\left (\frac{dI}{d\lambda} \right )_{ig} = 2x \\
    \left (\frac{dI}{d\lambda} \right )_{\rm tot} &= 7x \\
    V_{\rm tot} &= 3 V_{g} + V_{ig} = 0 + k B_{\rm true} x, 
\end{align}
where we assumed that the intensity gradient in the intergranular lanes is $x$, and that gradients can be added (as the wavelength derivative is a linear operation). The inferred magnetic field from this unresolved case will now be:
\begin{equation}
    kBx = V_{\rm tot} = k B_{\rm unresolved} 7 x, 
\end{equation}

where $B_{\rm true}$ is the magnetic field in the intergranular lane, equal to 1000\,G, but $B_{\rm unresolved} = {1000 {\rm G}}/{7} \approx 143$\,G, which is almost two times less than in the spatially resolved case. The reason for this discrepancy is again the nonlinearity of the inference method, which comes into play even in this simple approach. The M-E approximation behaves similarly to this case, because Stokes $V$ also depends linearly on the gradient of the source function, which is different in different observed pixels.

The effect of the telescope PSF on the estimated mean flux density has been pointed out by \citet{Plowman2}, where the authors ascribe this effect to the mixing of pixels with different velocities. To test the importance of this effect, we performed two additional tests. First, we repeated the whole experiment (spectral line synthesis and the M-E inversion for different levels of PSF degradation) using the same MHD cube but setting all the velocities to zero. This eliminates the line shifts and asymmetries altogether. The results for the mean flux density we obtain by inverting this dataset are essentially unchanged, suggesting that velocities play almost no role. Next, we set the velocities to zero and made each atmosphere have identical temperature and pressure stratification (but preserved the original magnetic field stratification in each pixel). This results in a thermodynamically homogeneous atmosphere in the $(x,y)$ plane, where pixels only differ in the magnetic field structure. In this case, the convolution with the telescope PSF affects the inferred mean magnetic flux density to a much lesser extent. The inversion resulted in mean flux densities of 32\,G for the case without spatial degradation and 28.9\,G for the PSF of a 10\,cm telescope. The identical temperature structure causes all pixels to have the same source function gradient, thus eliminating the above effect to a high degree. Given that the amplitudes of the Stokes $Q$, $U,$ and~$V$ intensities are proportional to the gradient of the source function with optical depth, this result is expected as the spatial blurring has no net effect in this parameter. In the WFA, this dependency is contained in the $dI_\nu/d\lambda$ term.

\subsection{Depth dependence of the magnetic field}
\label{depthtau}

\begin{figure}
\includegraphics[width=0.49\textwidth]{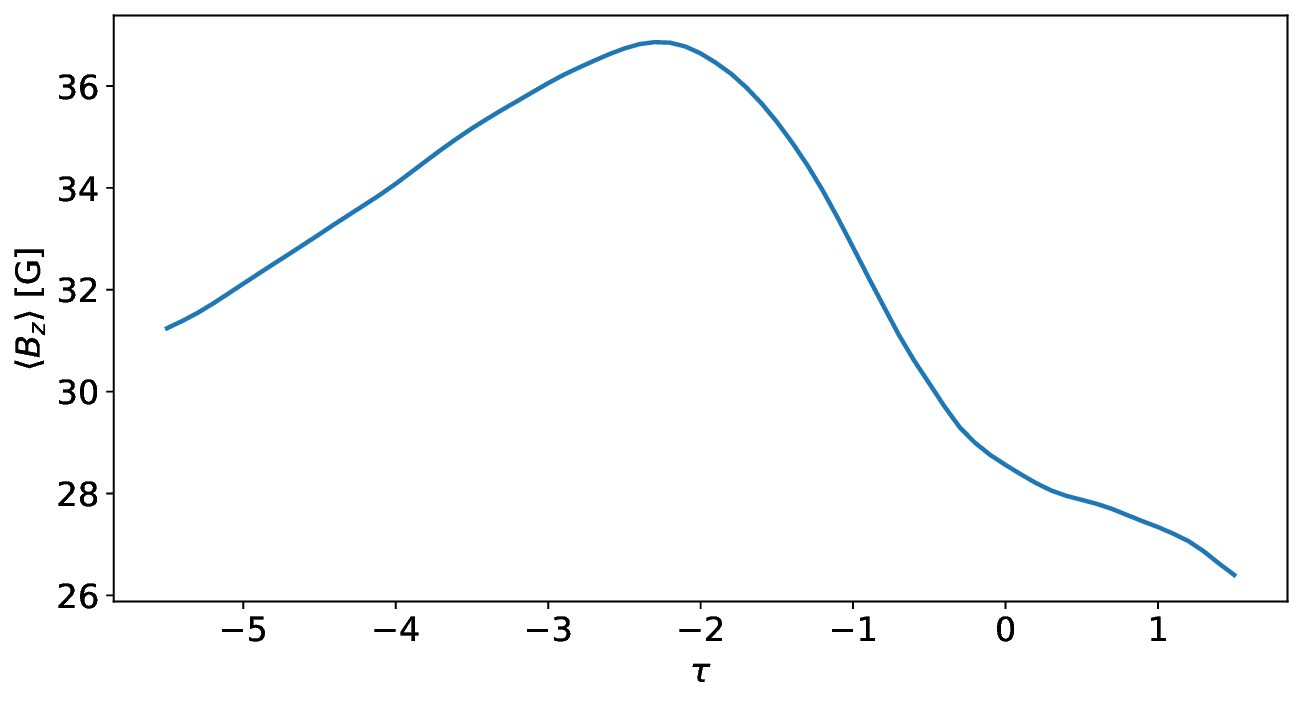}
\caption{Variation of the mean magnetic flux density with optical depth in the MURaM cube.}
\label{corr}
\end{figure}

A less pronounced but evident effect is that the inferred mean magnetic flux density ---even from the ideal, spatially undegraded noise-free spectra--- is $\sim$10\% higher than the original one. This has already been reported by \citet{Schlieche2023}. Their explanation is that the more strongly magnetic regions are also the less opaque ones, thus allowing us to see deeper into the atmosphere, where the magnetic field is even stronger. We note that, because of inhomogeneities in physical conditions in different pixels, we effectively probe the magnetic field at different depths. That is, we are inferring a magnetic field over a corrugated surface. To a first degree, this surface can be approximated by a $\log\tau=$~const surface. In the case of the Hinode lines, an estimate of the depth where the M-E inversion probes the magnetic field is, on average, approximately $\log\tau=-1.5$ \citep{WP1998, paperI}. To illustrate the effect of this corrugation, in Fig.\,\ref{corr} we plot the mean magnetic flux density versus the continuum optical depth in the MURaM simulation. The values range between 27\,G in the deep photosphere ($\log\tau =1)$ and 36\,G in the upper-mid photosphere ($\log\tau=-2$). This amounts to around $30\%$ difference, which depends on the opacity calculations used to convert geometrical height to optical depth. We note that the  vertically stratified inversions retrieve physical parameters on an optical depth grid, and so according to Fig.\,\ref{corr}, it is expected that depth-dependent inversions will not result in the conservation of the mean magnetic flux with optical depth. The only way to obtain a consistent solution is to retrieve the true height stratification. This was mostly attempted through post-processing of inversion results \citep[see: ][]{Puschmann2010, Loptien2018} but also by searching for a self-consistent MHD solution \citep{Tino2017}, and by using deep learning \citep{MLinversionOG}. Recently, \citet{JMBI, JMBII} demonstrated how to perform this latter process self-consistently with the spectropolarimetric inversion, but the method in its present form only works in highly magnetized photospheric regions.

\section{Test using SST/CRISP data}

\begin{figure*}
\centering
\includegraphics[width=0.85\textwidth]{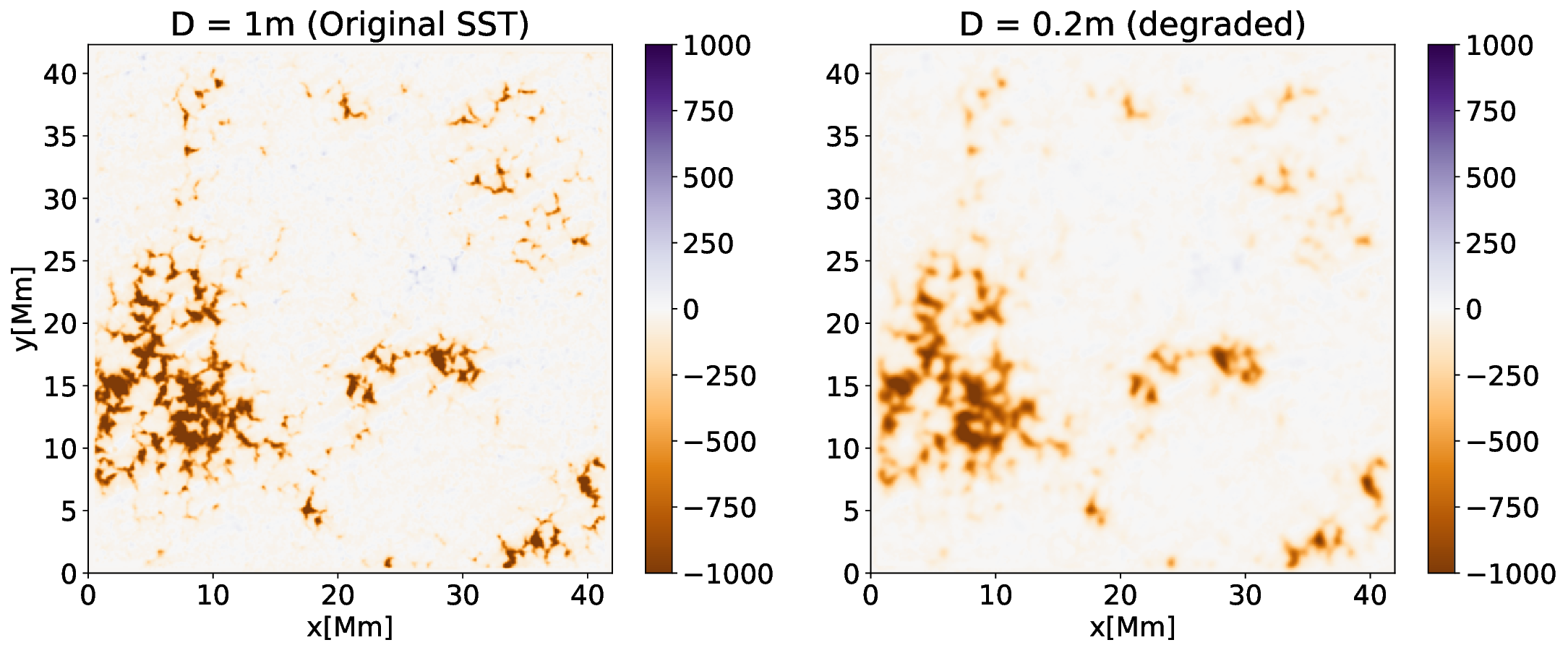}
\caption{LOS magnetic field from the original (left) and degraded (right) SST observations. Both plots are in units of Gauss. The mean flux densities for the two cases are $-80$\,G and $-68\,$G, respectively.}
\label{SSTdata}
\end{figure*}

Our results in the previous section indicate that a decrease in the telescope resolution results in a decrease in the estimated mean magnetic flux density in the observed region. We expect this to depend on the magnetic field distribution in the $(x,y)$ plane. To test our conclusions on real-life data, we use very high-spatial-resolution observations of a plage close to the disk center performed with the CRISP instrument \citep{CRISP} at the Swedish Solar Telescope \citep[SST, ][]{SST}. The data were processed using the CRISPRED/SSTRED data pipeline \citep{2015A&A...573A..40D,2021A&A...653A..68L} and atmospheric seeing effects were compensated using the Multi-Object-Multi-Frame-Blind-Deconvolution technique \citep{2002SPIE.4792..146L,2005SoPh..228..191V}. The observational setup is already described in \citet{Kianfarplage}, and so we summarize only the information relevant to our study. The FoV is situated very close to the disk center (-135'', 77'', $\mu=0.99$) and covers a 44 $\times$ 43 arcsec area, with a sampling of 0.059 arcseconds per pixel. The Fe\,I 6301/6302 spectral lines are sampled using a nonequidistant grid with a total of 16 wavelength points at 37\,s cadence. For the purposes of the experiment, we assumed that these observations fully resolve the field structure. We then convolved it with an Airy disk PSF corresponding to a 20\,cm aperture, and inverted both data sets using PyMilne, taking into account the CRISP spectral profile.  We calculated the mean vertical flux density and found it to be -80\,G for the original CRISP data and -68\,G for the degraded data (Fig\,\ref{SSTdata}). The effect exists but is noticeably lower: 15\%, compared to a 30\% loss of mean flux density found using the synthetic MURaM spectra. One difference is that the structuring of the magnetic field in the plage is different from the simulated magnetic field in the MURaM simulation. That is, more of the flux is contained in the lower spatial frequencies, making it less sensitive to the effects we find when analyzing the synthetic data. Another possibility is that a part of the flux density is already missing due to the PSF of the SST. That is, even high-resolution observations like these ones are still a degraded version of the ``real'' Sun. Overall, 15\% is a non-negligible amount and is probably comparable to or larger than the standard uncertainties in the magnetic field estimates. 

\section{Conclusions}

The main conclusion of our work is that even an estimate of a spatially averaged quantity ---in this case, the mean magnetic flux density $\langle B_z \rangle$--- suffers from a bias when the telescope resolution is insufficiently high. Here, we simulated this effect by calculating the spectra of magnetically sensitive lines of neutral iron around 630\,nm, applying various levels of spatial PSF, and inverting the data using a conventional, M-E model. This effect was previously found by \citet{Plowman2}, but it was attributed to the presence of velocities. Here, we show that the most important factor causing this bias is the nonlinearity in both the forward and inverse model used to infer the magnetic field. Another reason for a systematic error in the estimation of the mean flux density is the corrugation of iso-optical depth surfaces compared to iso-height surfaces. Even though these effects are rather simple to find and test for, the mechanisms causing them lie in the physics of the spectral line formation in the inhomogeneous atmosphere and are not trivial to correct. Synoptic observations, for example, imply relatively low spatial resolution in order to cover the whole solar disk. Coincidentally, these are the very same observations that are used to compute the open flux and are likely to suffer from the biases explored in our work. \citet{Linker2017} posed the open flux problem as a possible misdiagnosis of the photospheric magnetic fields, which leads to a disagreement between measured and extrapolated magnetic fields in the heliosphere. The influence of spatial resolution on magnetic field extrapolations has been demonstrated by, for example, \cite{deRosa2015}, who degraded Hinode/SOT data, and by \citet{Fleishman2017}, who directly rebinned magnetic fields from the numerical simulation. Our results directly identify some of the biases and confirm that some of the open flux in the photosphere might be lacking due to the limited telescope resolution. 

Overall, our findings are in support of using high-resolution observations to perform magnetic fields extrapolations \citep[e.g.,][]{Vissers2022}, and calibrating the low-resolution observations to the higher resolution ones \citep[e.g.,][]{HMIvsSOT}. To obtain a more complete understanding of the structure of the magnetic field, we must pursue and develop more sophisticated inversion techniques \citep[e.g.,][]{JMBI}. We will follow up this study with a more in-depth analysis of the biases found in the polar observations, following the conclusions made in \citet{paperI}, but focusing on the effects of PSF, noise, and disambiguation. A reliable quantitative inference of the open flux is especially critical in the scope of the upcoming observations of the solar poles by the Solar Orbiter mission. 

\begin{acknowledgement}
We thank Milan Go\v{s}i\'{c}, Bryan Yamashiro, and Marian Martin\'{e}z Gonzalez for fruitful discussions and ideas related to the research presented in this paper. IM thanks the solar magnetism group at KIS for their remarks on the manuscript. We thank an anonymous referee, whose comments helped improve the manuscript. 

IM and XS acknowledge support from NASA HGIO award 80NSSC21K0736. IM also acknowledges the International Space Science Institute (ISSI) in Bern, through ISSI International Team project 497 (Advanced Three-Dimensional Modeling of the Magnetic Field in Active Regions on the Sun).  RC acknowledges support from NASA LWS Award 80NSSC20K0217. This material is based upon work supported by the National Center for Atmospheric Research, which is a major facility sponsored by the National Science Foundation under Cooperative Agreement No. 1852977. We would like to acknowledge high-performance computing support from Cheyenne (doi:10.5065/D6RX99HX) provided by NCAR's Computational and Information Systems Laboratory, sponsored by the National Science Foundation.
This project has received funding from the European Research Council (ERC) under the European Union's Horizon 2020 research and innovation program (SUNMAG, grant agreement 759548). The Swedish 1-m Solar Telescope is operated on the island of La Palma by the Institute for Solar Physics of Stockholm University in the Spanish Observatorio del Roque de los Muchachos of the Instituto de Astrof\'isica de Canarias. The Institute for Solar Physics is supported by a grant for research infrastructures of national importance from the Swedish Research Council (registration number 2021-00169).
\end{acknowledgement}

\bibliography{psf_effects_final}


\end{document}